\pdfoutput=1

\documentclass[11pt, english]{article}
\usepackage{fullpage}
\usepackage{amsmath}
\usepackage{amssymb}
\usepackage{amsthm}
\usepackage{pxfonts}
\usepackage[T1]{fontenc}
\usepackage{color}
\usepackage{graphicx}
\usepackage{float}
\usepackage{hyperref}
\usepackage{charter}
\usepackage{algorithm}
\usepackage{algorithmic}

\newcommand{\EDF}{{\tt EDF}}
\newcommand{\EDFUS}{{\tt EDF-US[$1/2$]}}
\newcommand{\drop}[1]{#1}

\newcommand{\MPI}{{\tt MPI}}
\newcommand{\PVM}{{\tt PVM}}

\newcommand{\equals}{\stackrel{\mathrm{def}}{=}}
\newcommand{\lcm}{\mathrm{lcm}}
\newcommand{\bigOh}{\mathcal{O}}

\newtheorem{theorem}{Theorem}
\newtheorem{corollary}[theorem]{Corollary}
\newtheorem{definition}[theorem]{Definition}

\title{Integrating job parallelism\\
in real-time scheduling
theory\footnote{This is an extended version of the paper \emph{Integrating Job Parallelism in Real-Time Scheduling Theory}, published in the journal \emph{Information Processing Letters}. Ben Rodriguez$^\dagger$ pointed that the definition of work-limited parallelism did not express constraints that we were using implicitly in the remainder of the paper. Consequently we adapted the definition and clarified the proof.}}
\author{Sébastien Collette\thanks{Computer Science Department, Universit\'e Libre de Bruxelles, CP212, Bvd. du Triomphe, 1050 Brussels, Belgium}~~\thanks{Chargé de recherches du F.R.S.-FNRS, \protect\url{sebastien.collette@ulb.ac.be}. }
 \and Liliana Cucu\footnotemark[2]~~\thanks{\protect\url{liliana.cucu@ulb.ac.be}}
 \and Jo\a"el Goossens\footnotemark[2]~~\thanks{\protect\url{joel.goossens@ulb.ac.be}.}}
\date{}

\begin{document}
\maketitle

\begin{abstract}
We investigate the global scheduling of sporadic, implicit
deadline, real-time task systems on multiprocessor
platforms. We provide a task model which integrates job
parallelism. We prove that the time-complexity of the
feasibility problem of these systems is linear relatively to
the number of (sporadic) tasks for a fixed number of
processors. We propose a scheduling algorithm theoretically
optimal (i.e., preemptions and migrations neglected).
Moreover, we provide an \emph{exact} feasibility utilization
bound. Lastly, we propose a technique to limit the number of
migrations and preemptions.
\end{abstract}


\sloppy
\section{Introduction} \label{intro} 
The use of computers to control safety-critical real-time
functions has increased rapidly over the past few years. As a
consequence, real-time systems --- computer systems where the
correctness of each computation depends on both the logical results of
the computation and the time at which these results are produced ---
have become the focus of much study. Since the concept of ``time'' is
of such importance in real-time application systems, and since these
systems typically involve the sharing of one or more resources among
various contending processes, the concept of scheduling is integral to
real-time system design and analysis. Scheduling theory as it pertains
to a finite set of requests for resources is a well-researched
topic. However, requests in real-time environments are
often of a recurring nature. Such systems are typically modeled as finite
collections of simple, highly repetitive tasks, each of which
generates \emph{jobs} in a very predictable manner. These
tasks have bounds upon their worst-case execution requirements and their periods,
and associated deadlines. 

In this work, we consider {\it sporadic} task systems, i.e.,
where there is at least $T_i$ (called the period) time units
between two consecutive job instances of the same task. 
A job which occurs at time $t$ must be executed for
at most $C_i$ time units in the time interval
$[t,t+D_{i})$ (where $D_{i}$ is the relative deadline). A
particular case of sporadic tasks are the
\emph{periodic} ones for which the period is the
\emph{exact} temporal separation between the arrival of two
successive jobs generated by the task. We shall distinguish
between \emph{implicit deadline} systems where $D_i=T_i,
\forall i$; \emph{constrained deadline} systems where
$D_i \leq T_i, \forall i$; and \emph{arbitrary deadline}
systems where there is no constraint between the deadline
and the period. Moreover, we assume that the various tasks
are \emph{independent} (i.e., except the $m$ processors there
are no other shared resources, no critical sections nor
precedence constraints).

The \emph{scheduling algorithm} determines which job[s]
should be executed at each time instant. We distinguish
between \emph{off-line} and \emph{on-line} schedulers.
On-line schedulers construct the schedule during the
execution of the system; while off-line schedulers mimic
during the execution of the system a precomputed
(off-line) schedule. Remark that if a task is not active at a given
time instant and the off-line schedule planned to execute
that task on a processor, the latter is simply idled (or used
for a non-critical task).

When there is at least one schedule satisfying all constraints of the system, the system is said to be
\emph{feasible}. More formal definitions of these notions are given in
Section~\ref{model}.

\emph{Uniprocessor} sporadic (and periodic) real-time
systems are well studied since the seminal paper of {Liu}
and {Layland}~\cite{Liu} which introduces a model of
implicit deadline systems. For uniprocessor systems we
know that the worst case arrival pattern for sporadic tasks
corresponds to the one of (synchronous and) periodic tasks
(see, e.g.~\cite{thesisMok}). Consequently, most of the
results obtained for periodic tasks remain for sporadic ones. 
Unfortunately, this is not the case
upon multiprocessors due to scheduling anomalies (see,
e.g.~\cite{Andersson2003Static-priority}).

The literature considering scheduling algorithms and feasibility tests
for uniprocessor scheduling is tremendous. In contrast for
\emph{multiprocessor} parallel machines the problem of meeting timing
constraints is a relatively new research area.

\subsection{Model}
We deal with jobs which may be executed on different
processors at the very same instant, in which case we say
that \emph{job parallelism} is allowed. For a task $\tau_i$ and
$m$ identical processors we define a $m$-tuple of real
numbers $\Gamma_i\equals
(\gamma_{i,1},\ldots, \gamma_{i,m})$ with
the interpretation that a job of $\tau_i$ that executes for $t$ time
units on $j$ processors completes $\gamma_{i,j} \times t$ units of
execution. Full parallelism, which corresponds to the case
where $\Gamma_{i}=(1,2,\ldots,m)$ is not realistic;
moreover, if full parallelism is allowed the
multiprocessor scheduling problem is equivalent to the
\emph{uni}processor one (by considering, e.g., a unique processor
$m$ times faster).

In this work, we consider \emph{work-limited} job
parallelism with the following definition: 

\begin{definition} [work-limited parallelism]\label{wljp}
The job parallelism is said to be work-limited if and only if for all
$\Gamma_{i}$ we have:
$$\forall 1\le i \le n,\, \forall 1 \le j < j' \le m,\;$$
$$\frac{j'}{j} > \frac{\gamma_{i,j'}}{\gamma_{i,j}} \text{~~~and~~~}$$
$$\gamma_{i,(j'+1)}-\gamma_{i,j'}\le\gamma_{i,(j+1)}-\gamma_{i,j}$$
\end{definition}

\noindent Note that the last restriction is equivalent to
$$\frac{\gamma_{i,(j'+c)}-\gamma_{i,j'}}{c}\le\frac{\gamma_{i,(j+d)}-\gamma_{i,j}}{d} $$

For instance, the $m$-tuple $\Gamma_i=(1.0, 1.1, 1.2, 1.3,
\mathbf{4.9})$ is \emph{not} a work-limited job parallelism,
since $\gamma_{i,5} = \mathbf{4.9} > 1.3 \times \frac{5}{4} =
1.625$.
These restrictions may at first seem strong, but are in fact
intuitive: we require that parallelism cannot be achieved
for free, and that even if adding one processor decreases
the time to finish a parallel job, a parallel job on $j'$
processors will never run $j'/j$ times as fast as on $j$
processors. Moreover, if going from $j$ to $j+1$ processors implies some performance loss, then going from $j+1$ to $j+2$ processors must impact the performance by at least the same amount\footnote{Ben Rodriguez pointed out that we implicitly use this last restriction in the remainder of the paper, and that it is thus required in the definition of \emph{work-limited} parallelism.}.

 Many applications fit in this model, as the
increase of parallelism often requires more time to
synchronize and to exchange data between parallel
processes.
Remark that work-limited parallelism requires that for
each task (say $\tau_{i}$), the quantities $\gamma_{i,j}$
are distinct ($\gamma_{i,1}<\gamma_{i,2}<\gamma_{i,3}<\cdots$).

\subsection{Related research} 
Even if the multiprocessor scheduling
of sporadic task systems is a new research field, important results
have already been obtained. See, e.g.,~\cite{andersson,baker2,Baker2005An-analysis-of-,Srinivasan2002Deadline-based-,goossens3} for details.

All these works consider models of tasks where jobs use
\emph{at most} a single processor each time instant. This
restriction is natural for the uniprocessor scheduling since
only one processor is available at any time instant even if
we deal with parallel algorithms. Nowadays, the use of
parallel computing is growing (see, e.g.,~\cite{leiss});
moreover, parallel programs can be easily designed using the
Message Passing Interface
(\MPI~\cite{Gorlatch1998A-Generic-MPI-I,Lusk1999Using-MPI-:-por})
or the Parallel Virtual Machine
(\PVM~\cite{Sunderam1990PVM:-A-Framewor,Geist1994PVM:-Parallel-V})
paradigms. Even better, sequential programs can be
parallelized using tools like {\tt OpenMP}
(see~\cite{355074} for details). Therefore for the
multiprocessor case we should
be able to describe jobs that may be executed on different
processors at the same time instant. For instance, we find such requirements in
real-time applications such as robot arm
dynamics~\cite{zomaya96}, where the computation of dynamics
and the solution of a linear systems are both parallelizable
and contain real-time constraints.

Few models and results in the literature concern real-time
systems taking into account job parallelism.
\mbox{Manimaran} et al.\@ in~\cite{mani98} consider the
\emph{non-preemptive} \EDF{} scheduling of \emph{periodic}
tasks, moreover they consider \emph{moldable} tasks (the
actual number of used processors is determined before
starting the system and remains unchanged), while we
consider \emph{malleable} tasks (the number of assigned
processors to a task may change during the execution).
Meanwhile, their task model and parallelism restriction
(i.e., the sub-linear speedup) is quite similar to our model
and our parallelism restriction (work-limited). {Han} et
al.\@ in~\cite{han89} considered the scheduling of a
(finite) set of real-time jobs allowing job parallelism.
Their scheduling problem is quite different than our,
moreover they do not provide a real model to take into
account the parallelism. 

\subsection{This research}  In this paper, we deal with
\emph{global
  scheduling}\footnote{Job migration and preemption are allowed.} of
implicit deadline sporadic task systems with work-limited job
parallelism upon \emph{identical parallel machines}, i.e., where all
the processors are identical in the sense that they have the same
computing power. We consider the
feasibility problem of these systems, taking into account work-limited
job parallelism. For work-limited job parallelism we prove that the
time-complexity of the feasibility problem is linear relative
to the number of tasks for a
fixed number of processors. We provide a scheduling algorithm.

To the best of our knowledge there is no
such result in the literature and this manuscript provides a model,
a first feasibility test and a first exact utilization
bound for such kind of systems.

\subsection{Organization} This paper is organized as
follows. In Section~\ref{model}, we introduce our model of
computation. In Section~\ref{ComplexitySect}, we present the
main result for the feasibility problem of implicit deadline
sporadic task systems with work-limited job parallelism upon
identical parallel machines when global scheduling is used.
We prove that the time-complexity of the feasibility problem
is linear relative to the number of tasks when the number
of processors is fixed. We provide a linear scheduling
algorithm which is proved theoretically optimal and we give
an exact feasibility utilization bound. In
Section~\ref{sec:reduc}, we propose a technique to limit the
number of migrations and preemptions. We conclude and we
give some hints for future work in
Section~\ref{conclusions}.

\section{Definitions and assumptions}  \label{model}

 We consider the scheduling of sporadic task systems on $m$
identical processors $\{p_1, p_2, \ldots, p_m \}$. A task
system $\tau$ is composed of $n$ sporadic tasks $\tau_1,
\tau_2, \ldots, \tau_n$, each task is characterized by a
period (and implicit deadline) $T_i$, a worst-case execution
time $C_i$ and a $m$-tuple $\Gamma_i=(\gamma_{i,1},
\gamma_{i,2}, \ldots, \gamma_{i,m})$ to describe the job
parallelism. 

We assume that $\gamma_{i,0} \equals 0 \; (\forall i)$ in
the following. A job of a task can be scheduled at the very
same instant on different processors. In order to define the
degree of parallelization of each task $\tau_i$ we define
the execution ratios $\gamma_{i,j}, \forall j \in
\{1, 2,\ldots, m \}$ associated to each task-index of
processor pair. A job that executes for $t$ time units on
$j$ processors completes $\gamma_{i,j} \times t$ units of
execution.  In this paper we consider work-limited job
parallelism as given by Definition~\ref{wljp}.

We will use the notation $\tau_i\equals (C_i,T_i,\Gamma_i),
\forall i$ with $\Gamma_i=(\gamma_{i,1}, \gamma_{i,2},
\ldots, \gamma_{i,m})$ with $\gamma_{i,1} < \gamma_{i,2} <
\cdots< \gamma_{i,m}$. Such a sporadic task generates an
infinite sequence of jobs with $T_i$ being a lower bound on
the separation between two consecutive arrivals, having a
worst-case execution requirement of $C_{i}$ units, and an
implicit relative hard deadline $T_{i}$. We denote the
utilization of $\tau_i$ by $u_i
\equals \frac{C_i}{T_i}$. In our model, the period and the
worst-case execution time are integers. 

A task system $\tau$ is said to be {\it feasible} upon a
multiprocessor platform if under all
  possible scenarios of arrivals there exists at least one schedule in
  which all tasks meet their deadlines.

\subsection{Minimal required number of processors}  
Notice that a task $\tau_{i}$ requires more than $k$
processors simultaneously if $u_{i}>\gamma_{i,k}$; we denote
by $k_{i}$ the largest such $k$ (meaning that $k_{i}$ is the
smallest number of processor[s] such that the task
$\tau_{i}$ is schedulable on
$k_{i}+1$ processors):
\begin{equation}\label{eq:smalNbrProc}
  k_{i} \equals 
\begin{cases}  
0 & \text{if $u_{i}\le \gamma_{i,1}$}\\
\max_{k=1}^{m} \{k \mid \gamma_{i,k} < u_i\} &
\text{otherwise.}
\end{cases}
\end{equation}

For example, let us consider the task system $\tau=\{\tau_1,
\tau_2
\}$ to be scheduled on three processors. We have
$\tau_1=(6,4,
\Gamma_1)$ with $\Gamma_1=(1.0, 1.5, 2.0)$ and $\tau_2=(3,4,
\Gamma_2)$ with $\Gamma_2=(1.0, 1.2, 1.3)$. Notice that the
system is infeasible if job parallelism is not allowed
since $\tau_1$ will never meet its deadline unless it is
scheduled on at least two processors (i.e., $k_{1}=1$).
There is a feasible schedule if the task $\tau_1$ is
scheduled on two processors and $\tau_2$ on a third one
(i.e., $k_{2}=0$).

\subsection{Canonical schedule}

\begin{definition} [schedule $\sigma$]\label{defSched} 
For any task system $\tau = \{ \tau_1, \ldots, \tau_n \}$
and any set of $m$ processors $\{p_1, \ldots, p_m \}$ we
define the {\em schedule} $\sigma(t)$ of system $\tau $ at
instant $t$ as $\sigma : \mathbb{R}_{+}
\rightarrow \{0, 1, \ldots, n \}^m$ where $ \sigma(t)
\equals (\sigma_1(t), \sigma_2(t), \ldots,
\sigma_m(t))$ with

\hfill
\begin{equation*}
  \sigma_j(t) \equals \left\{
\begin{array}{ll}
0 \quad & 
\begin{minipage}{8cm}
if there is no task scheduled on $p_j$ at instant $t$
\end{minipage}\\[2ex]

i \quad & 
\begin{minipage}{8cm}
if $\tau_i$  is scheduled on $p_j$ at instant t
\end{minipage}
\end{array}
\right.
\end{equation*}
for all $1\le j \le m$.
\end{definition}\bigskip

We will now define \emph{canonical} schedules. In what follows we will prove that it is always possible to find such a canonical schedule for all feasible task systems (see Theorem~\ref{thmcanonical}). Refer to Figure~\ref{schedCan} for an example of a simple canonical schedule. Intuitively a schedule is canonical if it is a schedule where the tasks with higher indices are assigned to the highest available processors greadily.\smallskip

\begin{definition} [canonical schedule] \label{defSchedCan} For any task
  system $\tau = \{ \tau_1, \ldots, \tau_n \}$ and any set of $m$
  processors $\{p_1, \ldots, p_m \}$, a schedule $\sigma$ is
  \emph{canonical} if and only if the following
inequalities are satisfied:
 $$\forall 1\le j \le m, \,\forall 0\le t<t' < 1: \sigma_{j}(t') \le \sigma_{j}(t)$$
  $$\forall 1\le j<j'\le m, \,\forall t,t' \in [0,1):
\sigma_{j}(t) \le \sigma_{j'}(t')$$ 
\noindent and the schedule $\sigma$ contains a pattern that is
  repeated every unit of time, i.e., 
  $\forall t \in \mathbb{R}_{+},
  \,\forall 1\le j \le m:
\sigma_{j}(t)=\sigma_{j}(t+1)\text{.}$
\end{definition}\bigskip

Without loss of generality for the feasibility
problem, we consider a feasibility interval of length
$1$. Notice that the following results can be generalized to
consider any interval of length $\ell$, as long as $\ell$
divides entirely the period of every task.

\section{Our feasibility problem} \label{ComplexitySect}

In this section we prove that if a task system $\tau$ is feasible,
then there exists a canonical schedule in which all tasks meet their
deadlines. We give an algorithm which, given any task system,
constructs the canonical schedule or answers that no schedule
exists. The algorithm runs in $\bigOh(n)$ time with $n$ the number of tasks
in the system.

We start with a generic necessary condition
for schedulability using work-limited parallelism:\smallskip
\begin{theorem}\label{necessarycond}
In the work-limited parallelism model and using an off-line scheduling algorithm, a necessary condition for a sporadic task system $\tau$ to be feasible on $m$ processors is given by:
$$
\sum_{i=1}^{n}
\left(k_{i}+\frac{u_i-\gamma_{i,k_{i}}}{\gamma_{i,k_{i}+1}-%
\gamma_{i,k_{i}}} \right)\le m$$
\end{theorem}

\begin{proof}
As $\tau$ is feasible on $m$ processors, there exists a schedule $\sigma$ meeting every deadline. 
We consider any time interval $[t,t+P)$ with $P\equals \lcm{\{T_{1},T_{2},\ldots, T_{n}\}}$. 

Let $a_{i,j}$ denote the duration where jobs of a task
$\tau_{i}$ are assigned to $j$ processors on the interval
$[t,t+P)$ using the schedule $\sigma$. The sum $\sum_{j=1}^{m}
j\cdot a_{i,j}$ gives the total processor use of the task
$\tau_{i}$ on the interval (total number of time units for
which a processor has been assigned to $\tau_{i}$). As we
can use at most $m$ processors concurrently,  we know that

$$\sum_{i=1}^{n}\sum_{j=1}^{m} {j\cdot a_{i,j} }\le m \cdot P$$

\noindent otherwise the jobs are assigned to more than $m$ processors on the interval. If on some interval of length $\ell$, $\tau_{i}$ is assigned
to $j$ processors, we can achieve the same quantity of work on $j'>j$
processors on an interval of length $\ell \frac{\gamma_{i,j}}{\gamma_{i,j'}}$. In
the first case, the processor use of the task $i$ is $\ell\,j$, while in the second case it is 
$\ell j'\;\frac{\gamma_{i,j}}{\gamma_{i,j'}}$. By the first restriction that we
enforced on the tuple $\Gamma_{i}$ (see Definition~\ref{wljp}), we have
\begin{eqnarray*}
\hskip 2.3cm \ell j'\frac{\gamma_{i,j}}{\gamma_{i,j'}} & > & \ell j'\frac{\gamma_{i,j'}\frac{j}{j'}}{\gamma_{i,j'}}\\
&>&\ell j\\
\end{eqnarray*}
Let $\sigma'$ be a slightly modified schedule compared to $\sigma$, where $\forall i\not = i',\forall j, a'_{i,j}=a_{i,j}$. For the task $\tau_{i'}$, it is scheduled on $j'$ processors instead of $j<j'$ in $\sigma$ for some interval of length $\ell$, i.e.

$$a'_{i',j}=a_{i',j}-\ell$$
$$a'_{i',j'}=a_{i',j'}+\ell \frac{\gamma_{i,j}}{\gamma_{i,j'}}$$

Then, for that task $\tau_{i'}$, $$\sum_{j=1}^{m} j\cdot a'_{i',j}>\sum_{j=1}^{m} j\cdot a_{i',j}$$ \noindent  

This proves that increasing the parallelism yields an increased sum. It remains to prove that it is not better to increase the parallelism to $j'>j$ processors on some interval $\ell$ in order to decrease it to $j''<j$ processors on some other interval $\ell'$ (see Figure~\ref{newfig}).  

\begin{figure}[h]
\begin{center}
\includegraphics[scale=1]{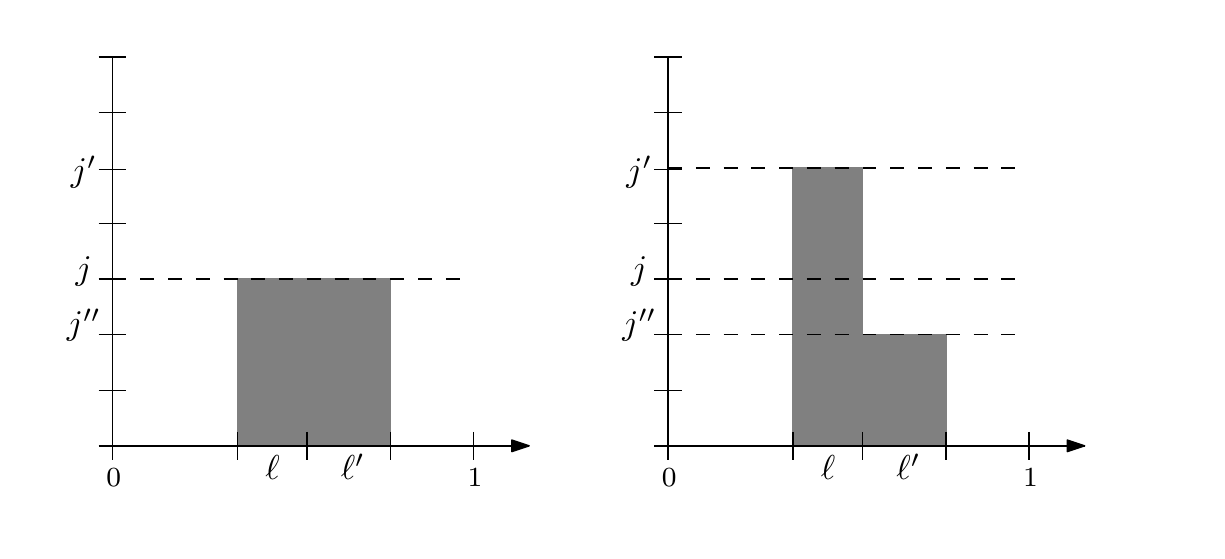}
\end{center}
\caption{Increasing the parallelism on some interval $\ell$ to be able to decrease it on $\ell'$.}
\label{newfig}
\end{figure}

The quantity of work originally achieved on the interval $\ell+\ell'$ is 
$(\ell+\ell')\cdot\gamma_{i,j}$ while the processor use is $(\ell+\ell')\cdot j$. After the change, the quantity of work is $\ell\gamma_{i,j'}+\ell'\gamma_{i,j''}$ for a processor use of $\ell j'+\ell' j''$. 

Suppose the processor use is not changed, and we shall show that the quantity of work has decreased. We start by noting that as the processor use is not changed, $(\ell+\ell')\cdot j=\ell j'+\ell' j''$, and we get 
\begin{eqnarray}
\ell (j'- j) = \ell' (j-j'')\label{eqn:pu}
\end{eqnarray}
 From the last restriction enforced by work-limited parallelism (see Definition~\ref{wljp}), we know that 
 \begin{eqnarray}\label{eqn:wl}
 \gamma_{i,j'}- \gamma_{i,j} \le \left(\gamma_{i,j}-\gamma_{i,j''}\right)\cdot\frac{j'-j}{j-j''}
 \end{eqnarray}

Now we have all the tools needed:
 \begin{eqnarray*}
 \ell\gamma_{i,j'}+\ell'\gamma_{i,j''}&=&\ell\left(\gamma_{i,j'}- \gamma_{i,j}\right) + \ell\gamma_{i,j}+\ell'\gamma_{i,j}-\ell'\left(\gamma_{i,j}-\gamma_{i,j''}\right)\\
&=&\ell\left(\gamma_{i,j'}- \gamma_{i,j}\right)-\ell'\left(\gamma_{i,j}-\gamma_{i,j''}\right)\frac{j-j''}{j-j''} + (\ell + \ell') \gamma_{i,j}\\
&=&\ell\left(\gamma_{i,j'}- \gamma_{i,j}\right)-\ell\left(\gamma_{i,j}-\gamma_{i,j''}\right)\frac{j'-j}{j-j''} + (\ell + \ell') \gamma_{i,j}~\text{(by Equation~\ref{eqn:pu})}\\
&\le&\ell\left(\gamma_{i,j}-\gamma_{i,j''}\right)\frac{j'-j}{j-j''}-\ell\left(\gamma_{i,j}-\gamma_{i,j''}\right)\frac{j'-j}{j-j''} + (\ell + \ell') \gamma_{i,j}~\text{(by Equation~\ref{eqn:wl})}\\
&\le&(\ell + \ell') \gamma_{i,j}\\
\end{eqnarray*}
In other words we decreased the quantity of work for a fixed amount of processor use; if we want to keep the same quantity of work, we need to increase the processor use and thus the sum defined above.

So, we proved that we should minimize the parallelism; as we want to derive a necessary condition, we schedule the task on the minimal number of processors required. 
A lower bound on the sum is then given by
$$k_{i}\cdot P+\frac{u_i-\gamma_{i,k_{i}}}{\gamma_{i,k_{i}+1}-\gamma_{i,k_{i}}}\cdot P$$ \noindent which corresponds to scheduling the task on $k_{i}+1$ processors for a minimal amount of time, and on $k_{i}$ processors for the rest of the interval. Then

$$\sum_{j=1}^{m} j\cdot a_{i,j} \ge  k_{i}\cdot P+\frac{u_i-\gamma_{i,k_{i}}}{\gamma_{i,k_{i}+1}-\gamma_{i,k_{i}}}\cdot P $$

and thus

\begin{eqnarray*}
\sum_{i=1}^{n}\sum_{j=1}^{m} {j\cdot a_{i,j} } &\le & m \cdot P\\
\hskip 0.7cm \sum_{i=1}^{n}\left(k_{i}\cdot P+\frac{u_i-\gamma_{i,k_{i}}}{\gamma_{i,k_{i}+1}-\gamma_{i,k_{i}}}\cdot P\right)  &\le & m \cdot P\\
\sum_{i=1}^{n}\left(k_{i}+\frac{u_i-\gamma_{i,k_{i}}}{\gamma_{i,k_{i}+1}-\gamma_{i,k_{i}}}\right)  &\le & m\\
\end{eqnarray*}
\noindent which is the claim of our theorem.
\end{proof}

\begin{theorem}\label{thmcanonical}
Given any feasible task system $\tau$,  the canonical
schedule meet all the deadlines.
\end{theorem}

\begin{proof}
  The proof consists of three parts: we first give an algorithm which
  constructs a schedule $\sigma$ for $\tau$, then we prove that
  $\sigma$ is canonical, and we finish by showing that the tasks
  meet their deadlines if $\tau$ is feasible.

  The algorithm works as follows (see Algorithm~\ref{algoSchedSteps}): 
  we consider sequentially every task
  $\tau_{i}$, with $i=n, n-1, \ldots ,1$ and define the schedule for
  these tasks in the time interval $[0,1)$, which is then repeated.

  We calculate the duration (time interval) for which a task $\tau_{i}$ uses $k_i+1$
  processors.  If we denote by $\ell_{i}$ the duration that the task $\tau_i$
  spends on $k_i+1$ processors, then we obtain the following equation:
  \begin{equation*}
    \label{eq:DurProcMax}
    \ell_{i}\,\gamma_{i,k_i+1}+(1-\ell_{i})\,\gamma_{i,k_i}=u_i
  \end{equation*}

  Therefore we assign a task $\tau_{i}$ to $k_{i}+1$ processors for a
  duration of $$\frac{u_{i}-\gamma_{i,k_{i}}}{\gamma_{i,k_{i}+1}-\gamma_{i,k_{i}}}$$
  and to $k_{i}$ processors for the remainder of the interval, which
  ensures that the task satisfies its deadline, since each
job generated by the sporadic task $\tau_{i}$ which arrives
at time $t$ receives in the time interval $[t,t+T_{i})$
exactly $T_{i}\times u_{i} = C_{i}$ time units. 

The task $\tau_{n}$ is
  assigned to the processors $(p_{m}, \ldots ,p_{m-k_n})$ (see
  Figure~\ref{schedCan}). If
  $u_{n}\not =\gamma_{n,k_{n}+1}$, another task can be scheduled at the end
  of the interval on the processor $p_{m-k_n}$, as $\tau_{n}$ does
  not require $k_{n}+1$ processors on the whole interval.

\begin{figure}[h]
\begin{center}
\includegraphics[scale=1.25]{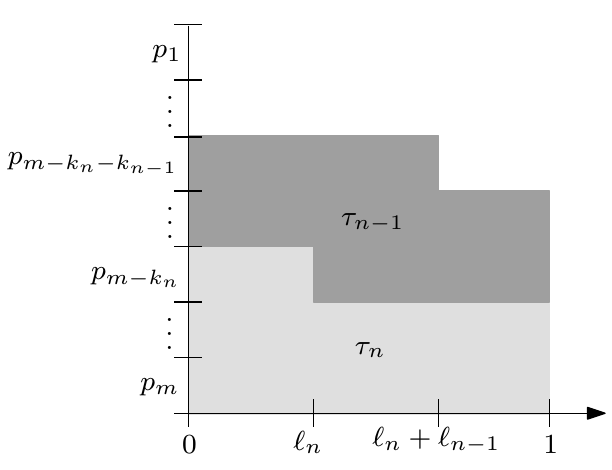}
\end{center}
\caption{Schedule obtained after scheduling the task $\tau_n$}
\label{schedCan}
\end{figure}

  We continue to assign greedily every task $\tau_{i}$, by first
  considering the processors with highest number.
The schedule produced by the above algorithm is canonical as it
respects the three constraints of the definition: 

\begin{itemize}
\item on every processor $j$ we assign tasks by decreasing index, thus
  $\sigma_{j}(t)$ is monotone and decreasing;
\item for all $i<i'$, if
$\tau_{i'}$ is scheduled on a processor $p_{j'}$, then $\tau_{i}$ is
assigned to a processor $p_{j}$ with $j\le j'$;
\item  the schedule
is repeated every unit of time.
\end{itemize}

The last step is to prove that if our algorithm fails to construct a
schedule, i.e., if at some point we run out of processors while there
are still tasks to assign, then the system is infeasible.

Let $\lambda_{i}$ be the  the total processor use of the task
$\tau_{i}$ in every unit length time interval. In the case of a canonical schedule, $\lambda_i$ corresponds to:
$$\lambda_{i} =
k_{i}+\frac{u_i-\gamma_{i,k_{i}}}{\gamma_{i,k_{i}+1}-\gamma_{i,k_{i}}}\text{.}$$ So for instance if, in the canonical schedule, a task $\tau_{i}$ is assigned to
$\lambda_{i}=2.75$ processors, it means that it is scheduled on two
processors for $0.25$ time unit in any time interval of length $1$,
and on three processors for $0.75$ time unit in the same interval.

If our algorithm fails, it means that $\sum_{i=1}^{n}{\lambda_{i}}>m$, which by Theorem~\ref{necessarycond} implies that the system is
infeasible.
\end{proof}

A detailed description of the scheduling
algorithm is given by Algorithm~\ref{algoSchedSteps}.
\begin{algorithm}
  \caption{{ Scheduling algorithm of implicit deadline sporadic
      task system $\tau$ of $n$ tasks on $m$ processors with
     work-limited job parallelism}}
  \label{algoSchedSteps}

  \begin{algorithmic}[1]
    \REQUIRE {The task system $\tau$ and the number
of processors $m$}

    \ENSURE { A canonical schedule of $\tau$ or a certificate that
      the system is infeasible}

    \STATE { let $j=m$} 
    \STATE { let $t_0=0$} 
    \STATE { let $\sigma_{p}(t)=0, \forall t \in [0,1), \forall 1\le p \le m$}
    \FOR{$i=n$ downto $1$}  
\IF{ $u_{i}\le \gamma_{i,1}$}
\STATE{ let $k_{i}=0$}
\ELSE
\STATE { let $k_{i} = \max_{k=1}^{m} \{k \mid \gamma_{i,k} < u_i\}$} 
\ENDIF
       \FOR{$r=1$ upto $k_{i}$}
         \STATE{ let $\sigma_{j}(t)=i, \forall t \in [t_{0},1)$}
         \STATE{ let $\sigma_{j-1}(t)=i, \forall t \in [0,t_{0})$}
         \STATE{ let $j=j-1$}
       \ENDFOR
      \STATE{ let $tmp=t_{0}+\frac{u_i-\gamma_{i,k_i}}{\gamma_{i,k_i+1}-\gamma_{i,k_i}}$}
      \IF{ $tmp>1$}

       \STATE{ let $\sigma_{j}(t)=i, \forall t \in [t_{0},1)$}
\STATE{ let $j=j-1$}
\STATE{ let $t_{0}=0$}
\STATE{ let $tmp=tmp-1$}
      \ENDIF
      \STATE{ let $\sigma_{j}(t)=i, \forall t \in [t_{0},tmp)$}
      \STATE{ let $t_{0}=tmp$} 
          \IF { $j \le 0$}
            \STATE { return Infeasible}
          \ENDIF 
    \ENDFOR
  \end{algorithmic}
  \end{algorithm}
If we consider the task system $\tau=\{\tau_1, \tau_2\}$ given
  before we have $k_1=1 $ and $k_2=0$. By using Algorithm~\ref{algoSchedSteps} we obtain: 
 
\begin{equation*}
 \begin{array}{l}
 \sigma_3(t)=2, \forall t \in [0, 0.75) \\
  \sigma_3(t)=1, \forall t \in [0.75, 1) \\
   \sigma_2(t)=1, \forall t \in [0, 1) \\
    \sigma_1(t)=1, \forall t \in [0, 0.75)\\
     \sigma_1(t)=0, \forall t \in [0.75, 1)\text{.} \\
 \end{array}
 \end{equation*}

Notice that Algorithm~\ref{algoSchedSteps} does not provide
satisfactory schedules for problems for which the number
of migrations and preemptions is an issue. We shall
address this question in the next section.\bigskip

\begin{corollary}\label{corollary::utilization-bond}
  In the work-limited parallelism model and using an off-line scheduling algorithm, a necessary  \textbf{and} sufficient condition for a sporadic task system $\tau$ to be feasible on $m$ processors is given by:
$$\sum_{i=1}^{n} \left(k_{i}+\frac{u_i-\gamma_{i,k_{i}}}{\gamma_{i,k_{i}+1}-\gamma_{i,k_{i}}} \right)\le m$$ 
\end{corollary}

Please notice that
Corollary~\ref{corollary::utilization-bond} can be seen as 
\emph{feasibility utilization bound} and in particular a
\emph{generalization} of the bound for uniprocessor
(see~\cite{Liu}) where a sporadic and implicit deadline task
system is feasible if and only if $\sum_{i=1}^n u_{i} \leq
1$. Like the \EDF{} optimality for sporadic implicit
deadline tasks is based on the fact that $\sum_{i=1}^n u_{i} \leq
1$ is a sufficient condition, we prove the optimality of
the canonical schedule based on the fact that $\sum_{i=1}^{n}
\left(k_{i}+\frac{u_i-\gamma_{i,k_{i}}}{\gamma_{i,k_{i}+1}-
\gamma_{i,k_{i}}} \right)\le m$ 
is a sufficient condition.\bigskip

\begin{corollary} \label{optComp}
  There exists an algorithm which, given any task system, constructs
  the canonical schedule or answers that no schedule exists
in $\bigOh(n)$
  time.
\end{corollary}
\begin{proof}
  We know that the algorithm exists as it was used in the proof of
  Theorem~\ref{thmcanonical}. For every task, we have to compute the
  number of processors required (in $\bigOh(1)$ time, as the number of
  processors $m$ is fixed), and for every corresponding processor
  $j$, define $\sigma_{j}(t)$ appropriately. In total, $\bigOh(n)$ time is
  required.
\end{proof}


\section{Scheduling problem reduction}
\label{sec:reduc}

\begin{figure}[htb]
\begin{center}
\includegraphics[width=7.5cm]{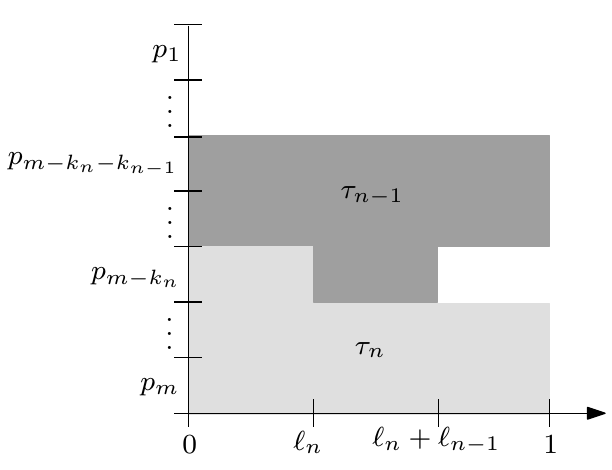}
\end{center}
\caption{Improved scheduling of $\tau_{n-1}$ and $\tau_n$}
\label{schedprime}
\end{figure}

Regarding optimality, we proved that each
task (say $\tau_{i}$) must use \emph{permanently} $k_{i}$
processor[s] \emph{simultaneously}, and that optionally
$\tau_{i}$ has to use an additional processor
\emph{partially} (i.e., with a duration \emph{strictly} less
than the unity in each interval of length 1). Regarding the
number of migrations, we can however define a \emph{better}
schedule without loss of optimality. For instance, in the
schedule given by Figure~\ref{schedCan}, task $\tau_{n-1}$
migrates between $k_{n-1}+1$ processors each time unit. I.e.,
$\tau_{n-1}$ uses $k_{n-1}$ processors in $[0,\ell_{n})$,
$k_{n-1}+1$ processors in $[\ell_{n}, \ell_{n}+\ell_{n-1}]$
and again $k_{n-1}$ processors in
$[\ell_{n}+\ell_{n-1},1)$, but \emph{not} the very same
processors as the ones used in $[0,\ell_{n}]$.
Consequently, there is necessarily a job migration of
$\tau_{n-1}$ each time unit. We can however assign $k_{n-1}$ processors
\emph{statically} and \emph{permanently} to $\tau_{n-1}$ and an
additional processor \emph{sporadically}
(for a duration of $\ell_{n-1} \cdot T_{n-1}$,
see~Figure~\ref{schedprime}). In terms of optimality we use
exactly the same number of processors but the number of
migrations is null (at least for $\tau_{n-1}$ in this
example). Since we assign statically and permanently tasks to
processors we can actually reduce our scheduling problem: 
the scheduling problem of the $n$ (original) sporadic tasks
upon $m$ processors with work-limited parallelism can be
reduced to a more studied and simpler scheduling problem:
the scheduling of $n'$ ($n' \leq n$) sporadic tasks upon
$m'$ ($m'\leq m$) processors where job parallelism can
be forbidden (without loss of optimality).

We shall now formalize the scheduling problem reduction. Let
$\{\tau_{1}, \tau_{2}, \ldots, \tau_{n}\}$ be our (original)
sporadic task set to schedule on $m$ processors. By the
definition of the quantity $k_{i}$
(Eq.~\ref{eq:smalNbrProc}) and the optimality of
Algorithm~\ref{schedCan} (Theorem~\ref{thmcanonical}), the
scheduling problem can be reduced to the scheduling of the
sporadic task set $\tau' \equals \{\tau'_{1}, \tau'_{2},
\ldots, \tau'_{n}\}$ with $C'_{i} \equals \ell_{i} \cdot
T_{i}$, $T'_{i} \equals T_{i}$ upon $m' \equals m -
\sum_{i=1}^{n} k_{i}$ where job parallelism can be
forbidden (without loss of optimality).

While our (canonical) schedules are optimal (the original
one and the improved one) for $\tau$, the number of
preemptions can be too large (there is actually, \emph{at
least} one preemption each time unit for the set of tasks
$\tau'$). For the systems where this is an issue,
since job parallelism can be forbidden for $\tau'$,
\emph{with} potential loss of schedulability, we can schedule
$\tau'$ using, for instance, global \EDF{} (see~\cite{goossens3}), global
Rate Monotonic (see~\cite{andersson}) or using a partition
scheme (see~\cite{Andersson2003Static-priority} for instance) in order
to reduce nicely the number of preemptions. We shall analyze
more precisely the case of using \EDF{} to schedule
the task sub-set $\tau'$ in the next section.

\subsection{Scheduling $\tau'$ using \EDF}
In order to reduce strongly the number of preemptions in the
schedule of task sub-set $\tau'$ we can use \EDF{} (or one
of its variants) instead of the canonical schedule. We know
that the number of preemptions using \EDF{} is bounded from
above by the number of jobs in the set (and consequently,
the total number of \emph{context switches\/} is bounded by
twice the number of jobs). It can similarly be shown that
the total number of {\em interprocessor migrations\/} of
individual jobs is bounded from above by the number of jobs.
We know that the task sub-set $\tau'$ is schedulable using
the optimal canonical schedule since the following necessary
and sufficient condition is satisfied: 

$$\sum_{\tau'_{i} \in \tau'} u'_{i} \leq m'$$ 

Unfortunately, \EDF{} is not optimal in terms of the number of processors
required and the condition above is only necessary. Using,
e.g., \EDFUS{} (which gives top
priority to jobs of tasks with utilizations above $1/2$ and
schedules the remaining jobs using \EDF)
Baker~\cite{Baker2005An-analysis-of-} proved correct the
following \emph{sufficient} (not exact) feasibility
test: the task sub-set $\tau'$ is schedulable by \EDFUS{}
upon $\widehat{m}$ processors if

$$(2 \cdot\sum_{\tau'_{i} \in \tau'} u'_{i} - 1)  \leq
\widehat{m}$$

The trade-off is actually the
additional number of processors required: $
\left\lceil 2 \cdot \sum_{\tau'_{i} \in \tau'} u'_{i} - 1
\right\rceil - m'$.

\section{Discussions} \label{conclusions}

\drop{
\subsection{Job parallelism vs.\ task parallelism}
In this manuscript we study multiprocessor systems where
job parallelism is allowed. We would like to distinguish
between two kinds of parallelism, but first the
definitions: \emph{task parallelism} allows each task to
be executed on several processors at the same
time, while \emph{job parallelism} allows each job to
be executed on several processors at the same
time. If we consider constrained (or implicit) deadline
systems task parallelism is not possible. For
\emph{arbitrary} deadline systems, where several jobs of
the same task can be active at the same time, the
distinction makes sense. Task parallelism
allows the various active jobs of the same task to be
executed on a different (but unique) processor while job
parallelism allows each jobs to be executed on several
processors at the same time.
}

\drop{\subsection{Optimality and future work}} In this paper we study the
feasibility problem of implicit deadline sporadic task systems with
work-limited job parallelism upon identical parallel machines when
global scheduling is used. We prove that the
time-complexity of our
problem is linear relative to the number of tasks. We provide an optimal scheduling
algorithm that runs in $\bigOh(n)$ time and we give an exact
feasibility utilization bound.

Our algorithm is optimal in terms of the number of
processors used. It is left open whether an
optimal algorithm in terms of the number of preemptions
can be designed. As
a first step, we used an interval of length 1 to study the
feasibility problem. Without loss of optimality, we can
improve our algorithm to work on an interval of length equal
to the $\mathit{gcd}$ of the periods of every task, which
decreases the number of preemptions and migrations. We do not know,
however, if this improvement \emph{minimizes} the number of
preemptions and migrations.

The definition of work-limited job parallelism was given
here for identical processors. One should investigate an
extension of this definition to heterogeneous platforms. 

\section*{Acknowledgments}
The authors thank anonymous referees, whose remarks led us
to better present our results. We also thank Ben Rodriguez for interesting discussions on potential extensions of this work, as well as for key ideas to fix the proof of Theorem~\ref{necessarycond}.

\bibliographystyle{elsart-num-sort}
\bibliography{SporadicWorkLimitedIPL}

\begin{thebibliography}{10}
\expandafter\ifx\csname url\endcsname\relax
  \def\url#1{\texttt{#1}}\fi
\expandafter\ifx\csname urlprefix\endcsname\relax\def\urlprefix{URL }\fi

\bibitem{Andersson2003Static-priority}
B.~Andersson, Static - priority scheduling on multiprocessors, Ph.D. thesis,
  Chalmers University of Technology, Göteborg, Sweden (2003).

\bibitem{andersson}
B.~Andersson, S.~K. Baruah, J.~Jonsson, Static-priority scheduling on
  multiprocessors, Proceedings of the 22nd IEEE Real-time Systems Symposium
  (2001) 193--202.

\bibitem{Baker2005An-analysis-of-}
T.~P. Baker, An analysis of {EDF} scheduling on a multiprocessor, IEEE Trans.
  on Parallel and Distributed Systems 15~(8) (2005) 760--768.

\bibitem{baker2}
T.~P. Baker, S.~K. Baruah, Handbook of Real-Time and Embedded Systems, chap.
  Schedulability Analysis of Multiprocessor Sporadic Task Systems, Chapman and
  Hall, 2007, pp. 3--1 -- 3--15.

\bibitem{355074}
R.~Chandra, L.~Dagum, D.~Kohr, D.~Maydan, J.~McDonald, R.~Menon, Parallel
  programming in OpenMP, Morgan Kaufmann Publishers Inc., San Francisco, CA,
  USA, 2001.

\bibitem{Geist1994PVM:-Parallel-V}
A.~Geist, A.~Beguelin, J.~Dongarra, W.~Jiang, R.~Manchek, V.~Sunderam, {PVM}:
  Parallel Virtual Machine A Users' Guide and Tutorial for Networked Parallel
  Computing, MIT Press, 1994.

\bibitem{goossens3}
J.~Goossens, S.~Funk, S.~K. Baruah, Priority-driven scheduling of periodic task
  systems on multiprocessors, Real-time Systems 25 (2--3) (2003) 187--205.

\bibitem{Gorlatch1998A-Generic-MPI-I}
S.~Gorlatch, H.~Bischof, A generic {MPI} implementation for a data-parallel
  skeleton: Formal derivation and application to {FFT}, Parallel Processing
  Letters 8~(4) (1998) 447--458.

\bibitem{Lusk1999Using-MPI-:-por}
W.~Gropp (ed.), Using MPI: portable parallel programming with the
  message-passing interface, 2nd ed., Cambridge, MIT Press, 1999.

\bibitem{han89}
C.~Han, K.-J. Lin, Scheduling parallelizable jobs on multiprocessors,
  Proceedings of the 10th IEEE Real-Time Systems Symposium ({RTSS}'89) (1989)
  59--67.

\bibitem{leiss}
E.~L. Leiss, Parallel and Vector Computing, McGraw-Hill, Inc., 1995.

\bibitem{Liu}
C.~Liu, J.~Layland, Scheduling algorithms for multiprogramming in a
  hard-real-time environment, Journal of the ACM 20~(1) (1973) 46--61.

\bibitem{mani98}
G.~Manimaran, C.~Siva Ram~Murthy, K.~Ramamritham, A new approach for scheduling
  of parallelizable tasks in real-time multiprocessor systems, Real-{Time}
  Systems 15 (1998) 39--60.

\bibitem{thesisMok}
A.~Mok, Fundamental design problems of distributed systems for the
  hard-real-time environment, Ph.D. thesis, Laboratory for Computer Science,
  Massachusetts Institute of Technology (1983).

\bibitem{Srinivasan2002Deadline-based-}
A.~Srinivasan, S.~Baruah, Deadline-based scheduling of periodic task systems on
  multiprocessors, Information Processing Letters 84 (2002) 93--98.

\bibitem{Sunderam1990PVM:-A-Framewor}
V.~Sunderam, {PVM}: A framework for parallel distributed computing,
  Concurrency: Practice and Experience 2~(4) (1990) 315--339.

\bibitem{zomaya96}
A.~Y. Zomaya, Parallel processing for real-time simulation: A case study, IEEE
  Parallel and Distributed Technology: System and Technology 4~(2) (1996)
  49--55.

\end{thebibliography}
\end{document}